\documentclass[aps,showpacs,nofootinbib,preprintnumbers,amsmath,amssymb,superscriptaddress]{revtex4}

\usepackage{graphicx}

\begin{document}

\title{On the nature of thermal QCD phase transitions}

\author{Yoshitaka Hatta} 
\affiliation{Department of Physics, Kyoto University, Kyoto 606-8502, Japan}
\affiliation{The Institute of Physical and Chemical Research (RIKEN)
Wako, Saitama 351-0198, Japan}
\affiliation{High Energy Theory, Department of Physics, Brookhaven National Laboratory, Upton, NY 11973, U.S.A.}

\author{Kenji Fukushima}
\affiliation{Center for Theoretical Physics, Massachusetts Institute of Technology, Cambridge, MA 02139, U.S.A.}
\affiliation{Department of Physics, University of Tokyo, 7-3-1 Hongo, Bunkyo-ku,  Tokyo 113-0033, Japan}


\begin{abstract}
We argue that the chiral phase transition and the deconfinement phase transition in finite temperature QCD are just different limits of a single crossover phase transition driven by the lightest scalar field which is a mixture of the glueball and the sigma meson. 
The key mechanism, the level repulsion between the sigma and the glueball, gives not only a unified description of the finite temperature phase transitions, but also a consistent characterization of the entire QCD phase diagram in the temperature-quark mass plane.

\end{abstract}

\date{\today}

\preprint{MIT-CTP-3448}
\maketitle

\section{Introduction}
Soon after the discovery of asymptotic freedom, it has become common knowledge that Quantum Chromodynamics (QCD) under extreme conditions, such as high temperature and/or density, is a theory of weakly coupled quarks and gluons \cite{collins,parisi}. This phase of QCD, called the Quark-Gluon Plasma (QGP), has attracted a considerable attention over the past few decades.

Theoretically, it is imperative to establish the thermodynamic properties of QCD at finite temperature and density in order to confront the experimental data of the heavy-ion collision. The first principle lattice QCD calculations have revealed a great deal of information on the finite temperature phase transition(s) between the hadronic phase and the QGP. An important task would be to correctly interpret in concrete physical terms what is going on during the rapid transition.

In QCD, at finite temperature, it is well known that there are two different phase transitions in different parameter regions; the chiral phase transition in theories with massless quarks and the deconfinement phase transition in theories with infinitely heavy quarks. 
Since the parameters of the real world are rather close to the chiral limit, chiral symmetry is often emphasized in constructing phenomenological models and describing the phase transition. One could argue that the deconfinement phase transition would have no definite meaning in this parameter region and could be forgotten altogether.

Yet the word `deconfinement' is quite appealing. The QGP is often referred to as a deconfined phase. It is indeed tempting to interpret the rapid increase of the energy density observed in the lattice simulations as the liberation of the QCD degrees of freedom, or `deconfinement'. 

Perhaps the reason of adherence to deconfinement or the deconfinement phase transition is not merely an intuitive one. Since the early days of finite temperature QCD lattice simulations, it has been empirically known that the chiral phase transition and the deconfinement phase transition are tightly correlated although there seems to be no logical relation between them {\it a priori} \cite{Kogut,fukugita,Karsch,aoki,karsch2,allton,gat,toublan}. From these results one expects that some remnants of the deconfinement phase transition might be taken over by the chiral phase transition. 

In this paper, we argue that the above expectation is correct. The chiral phase transition and the deconfinement phase transitions are just different limits of a single crossover phase transition. The key observations are the glueball-sigma mixing and the level repulsion which `lock' the two phase transitions \cite{link}. This helps dissolve the ambiguous issue of `deconfinement' and enables us to have a clearer picture of the nature of the finite temperature QCD phase transition.

The paper is organized as follows. In Section II, we briefly review some basic facts about the chiral and deconfinement phase transitions and pose a `puzzle' long known in finite temperature lattice QCD simulations. Section III contains our main results. We present a possible scenario to explain the puzzle and argue that the phenomenon is a global feature of the QCD phase diagram in the ($T, m$) plane ($T$ is the temperature and $m$ is the current quark mass). In Section IV, we speculate about QCD-like theories such as QCD with adjoint quarks and in the large $N_{\rm c}$ limit. Section V is devoted to our conclusions.

\section{QCD phase transitions: a puzzle} 

The deconfinement phase transition is well defined only in theories with
infinitely heavy quarks.  Above the critical temperature, the Polyakov loop \cite{polyakov,sus,r},
\begin{eqnarray}
 L(\boldsymbol{x})=\mbox{tr} P\exp{\Bigl( -{\rm i}\int^{1/T}_0 t^a A_0^a(\boldsymbol{x},\tau){\rm d}\tau \Bigr) },\label{11}
\end{eqnarray}
where $P$ is the path ordering operator and $\{ t^a,\ a=1,2,\cdots ,N_{\rm c}^2-1\}$ are the SU($N_{\rm c})$ generators in the fundamental representation, develops a nonzero expectation value which breaks center symmetry. In an SU(3) pure gauge theory the
transition is known to be first order \cite{yaffe}.
When heavy quarks are added to the theory, the first order
phase transition weakens. It eventually ends in a second order phase transition
point \textsf{D} where $\langle L(\boldsymbol{x})\rangle$ becomes
continuous.  The universality class of this phase transition is that of the three dimensional Ising model.
The critical
quark mass at \textsf{D} is estimated to be $\sim{\rm O}(1)$ GeV
in the two-flavor lattice simulations \cite{ha,attig}.

In contrast, chiral symmetry is a true symmetry only in theories with massless quarks. The order parameter of the chiral phase transition is the quark condensate $\langle \bar{q}q\rangle$.
For three flavors, the phase transition is first order
\cite{snow,rob}. At finite but small quark mass (which we take to be
equal for all flavors), the first order transition persists and eventually 
ends in a second order phase transition point \textsf{C}. The jump in the chiral condensate disappears and the scalar meson field $\sigma\sim \bar{q}q$ (with certain flavor content) becomes gapless at this point \cite{Gavin,schmidt}.
The location of \textsf{C} and \textsf{D} depends on the number of flavors and their mass ratios. [For example, \textsf{C} lies on the $m=0$ axis for two flavors.] The argument in the next section does not depend on the details of the flavor sector, however.

\begin{figure}
\includegraphics{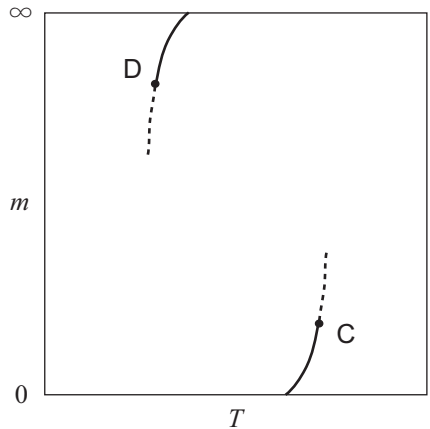}
  \caption{\label{2}  Naive phase diagram in the ($T$, $m$) plane. The solid (dotted) lines denote the first order (crossover) phase transition.}
\end{figure}

From these basic facts alone, obviously there is no logical relation between the chiral and deconfinement phase transitions. Various possibilities have been argued in the literature \cite{shuryak,rob2,larry}. A common argument is that, since confined vacuum necessarily breaks chiral symmetry \cite{casher,thooft,coleman}, which suggests that the energy scale of the chiral symmetry breaking is higher than that of confinement, chiral symmetry restoration occurs at a somewhat higher temperature than deconfinement. [See, however, \cite{ito,bar}.] This is in accord with our physical intuition; it is easy to imagine massive constituent quarks but it is hard to imagine massless protons. If there were no lattice simulations at finite temperature, theorists would have drawn a phase diagram like Fig.~\ref{2}, incorporating what is known from the universality argument.

\begin{figure}
\includegraphics{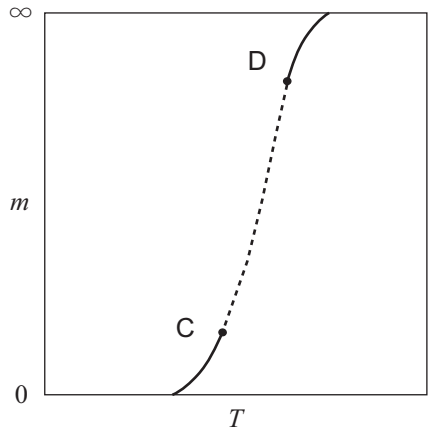}
  \caption{\label{fig} Realistic phase diagram observed in lattice simulations.}
\end{figure}

However, the realistic phase diagram  obtained from the lattice simulations, Fig.~\ref{fig}, is qualitatively different from our naive expectation \cite{Kogut,fukugita,Karsch,aoki,karsch2,gat,allton,toublan}. First, the critical temperature of the deconfinement phase transition in the pure gauge theory $\sim 280$ MeV is {\it invariably higher} than that of the chiral phase transition  in the chiral limit $\sim 170$ MeV. [One may think that it is meaningless to compare these two numbers because they are defined in completely different theories. However, this is a suggestive result and can be naturally explained within our scenario.] The more striking fact is that the pseudocritical temperatures determined from the peak position of the susceptibility of each order parameter (dotted curves in Fig.~\ref{fig}) appear to coincide for {\it all} values of the quark mass, which means that two crossover
curves are smoothly connected at an intermediate point.
It is as if the deconfinement phase transition in the pure gluonic theory is `guided' towards the chiral phase transition traveling over an infinite range of the quark mass. This is a surprising result. How can the deconfinement phase transition defined in a theory {\it without} quarks `know' about the chiral phase transition?

\section{Understanding the phase diagram}

The interplay between the two phase transitions is an interesting problem.
First Gocksch and Ogilvie have observed that what is responsible for the breaking of center symmetry is (the inverse of) the {\it constituent} quark mass rather than the current quark mass \cite{gock}. This suggests that the Polyakov loop and the sigma meson are closely coupled and the critical fluctuation of one field induces a singularity in the other field. More recent works \cite{satz,fuku,sa,fuku2} further confirmed this interplay. However, as pointed out in \cite{link} and emphasized in \cite{fuku2}, this argument alone is insufficient to explain the observed data. In fact, a stronger condition is needed to reproduce the phase diagram Fig.~\ref{fig}. In this section, we will make a serious attempt to understand the physics behind Fig.~\ref{fig}. We shall divide our argument into three steps. 

\subsection{Step 1: Massless glueball}

 Although there is little doubt that the Polyakov loop is the correct order parameter of the deconfinement phase transition in pure gauge theories, a perennial question about the physical meaning of the Polyakov loop remains \cite{smilga}. In a word, the Polyakov loop is a fictitious excitation in Euclidean space. It is true that, even in the presence of dynamical quarks, the behavior of the Polyakov loop indicates a clear signal of the phase change. 
However, it seems that the above conceptual barrier has long been an obstacle to our satisfactory understanding of the phase transition in the presence of dynamical quarks. Our first step is to propose an alternative characterization of the deconfinement phase transition. Specifically, we wish to show that the {\it glueball} can equivalently describe the critical phenomena at \textsf{D}.

For this purpose, first we work in the pure SU(2) gauge theory. In this case the deconfinement phase transition is second order \cite{la,ku}
whose universality class is that of the three dimensional
Ising model. The string tension (`mass' of the Polyakov loop) $K$ goes to zero smoothly as the critical temperature is approached
\begin{eqnarray}
K \sim |t|^{\nu}, \qquad (t<0) \label{sigma}
\end{eqnarray}
where $t$ is the reduced temperature and the universal exponent is $\nu \simeq 0.63$ \cite{Engels}. 

The infinite correlation of the Polyakov loop entails a
 massless mode in physical excitations.
This can be seen by considering the specific heat $C_{\rm v}$. Near the critical point, the singular part of the specific heat may be represented by
\begin{eqnarray}
 C_{\rm v} \sim \int {\rm d}^d x \langle
 L^2(\boldsymbol{x})L^2(\boldsymbol{0})\rangle_{\rm c}  \sim |t|^{-\alpha}, \label{c}
\end{eqnarray}
where $\alpha\simeq 0.11$ and $d=3$ is the spatial dimension \footnote{This is the standard expression of the specific heat in the field theoretic approach to critical phenomena \cite{zinn}. It is easily understood by noting that the temperature enters linearly in the `mass' term $\propto L^2$ of the Ginzburg-Landau model.}. The subscript c denotes the connected part \footnote{Note that $L(\boldsymbol{x})$ is real for SU(2) and $\langle L^2(\boldsymbol{x}) \rangle$ stays nonzero at any temperature.}.
In the theory of critical phenomena, the original dimension of an operator has no meaning. The Polyakov loop is dimensionless, but it has nothing to do with the actual {\it scaling dimension} which enters in the calculation of correlation functions. The scaling dimension of the order parameter squared is $d-1/\nu$ \cite{zinn}. Thus the explicit structure of the divergence is 
\begin{eqnarray}
 \int {\rm d}^d x \frac{{\rm e}^{-m_{\rm G}|\boldsymbol{x}|}}{|\boldsymbol{x}|^{2(d-\frac{1}{\nu})}}\sim |t|^{-\alpha}, \label{c2}
\end{eqnarray}
where we have assumed that the connected part of the correlator in
(\ref{c}) exponentially decays as $\sim {\rm e}^{-m_{\rm G}|\boldsymbol{x}|}$
at large distances.
Matching the powers of $|t|$ and using Josephson's law $\alpha=2-d\nu$ (hyperscaling), we obtain,
\begin{eqnarray}
m_{\rm G}\sim |t|^{\nu}. \label{ttt}
\end{eqnarray}
Hence the operator $L^2$ excites, among other states, at least one state with zero screening mass at the phase transition. 
By using an SU(2) identity 
\begin{eqnarray}
L^2(\boldsymbol{x})=1+L_{\rm A}(\boldsymbol{x}),
\end{eqnarray}
where $L_{\rm A}$ is the Polyakov loop in the adjoint representation (i.e., the generators $\{t^a\}$ in Eq.~(\ref{11}) are replaced by those in the adjoint representation), or just by 
noticing that $L^2$ is invariant under the center transformation, $L
\to -L$, we see that the massless state is the $0^{+}$ glueball
\footnote{Strictly speaking, the state created by the operator $L_{\rm A}$ should be called the `gluelump' instead of the glueball \cite{michael}. However, we expect that the difference (if any) dissolves at the deconfinement phase transition point. On the lattice, the specific heat is measured from the integrated correlator of plaquettes which excite glueballs. Since the two expressions of the specific heat should give the correct singularity, the gluelump and the glueball must be the same excitation near the phase transition.}.
Eq. (\ref{ttt}) is a trivial consequence of the scaling property at general second order phase transition points; the only relevant scale is the correlation length $\xi \sim |t|^{-\nu}$. Nevertheless, in the present case it is a nontrivial relation ensuring the vanishing of the mass gap in physical excitations as opposed to the Polyakov loop which has no particle interpretation in Minkowski space-time. Note that although the Polyakov loop and the glueball become massless with the same critical exponent $\nu$, their susceptibilities diverge with different critical exponents; $\alpha \sim 0.1$ for the glueball and $\gamma \sim 1.2$ for the Polyakov loop (the order parameter). [Compare with the SU(3) case below.] 
A significant decrease of the $0^{+}$ glueball screening mass near the
SU(2) deconfinement transition has been observed on the lattice \cite{datta}. \\

One might suspect that the vanishing of the glueball mass would contradict with the presence of the trace anomaly. In the simple phenomenological potential of glueballs at zero temperature \cite{sch,camp}, the glueballs acquire their mass entirely from the trace anomaly, or the gluon condensate. 
\begin{eqnarray}
m_{\rm G} \propto \bigl( \langle 0| \alpha_{\rm s} F_{\mu \nu}F^{\mu \nu}|0 \rangle \bigr) ^{\frac{1}{4}} \label{glu}
\end{eqnarray}
Since we do expect that the trace anomaly persists at any temperature, this appears to contradict with our assertion. 
However, the foundation of this objection is obscure.
First, the exact lattice sum rule for the glueball mass \cite{rothe} shows that only $\frac{1}{4}$ of the glueball mass comes from the expectation value of the trace anomaly in the glueball state measured {\it relative} to the vacuum.   
 Although it might be useful for certain purposes, Eq.~(\ref{glu}) cannot be taken seriously in our context.
Secondly, as observed in \cite{datta}, the symmetry organizing the glueball spectrum changes near the critical temperature of both the SU(2) and SU(3) pure gauge theories. At low temperatures, the zero temperature symmetry is dynamically realized in the glueball spectrum \cite{gross}. Near the phase transition, the zero temperature symmetry is superseded by the finite temperature one where the inclusion of the timelike links in interpolating operators makes a drastic change in the screening masses. The glueballs which become massless at the phase transition are the `electric' glueballs (those containing $A_0$, or time-like links), while the `magnetic' glueballs (composed of $A_i$, or space-like links) remain massive \footnote{Accordingly, we expect that the divergence of the specific heat comes from the correlator of timelike plaquettes.}. 
This crucial difference has been overlooked in most phenomenological approaches. [See, however, \cite{adrian}.]
Finally, but most cogently, Eq.~(\ref{ttt}) is true for any second order phase transition. If all the glueballs were massive at the phase transition, the specific heat would be finite \footnote{Due to the very small critical exponent, $\alpha \sim 0.1$, the correct singularity in the specific heat has not been confirmed on the lattice so far. See, however, \cite{eng}.}. This is a catastrophe because it means that the apparent singularity in the Polyakov loop has nothing to do with thermodynamics and there is no (second order) phase transition. \\

Let us go back to the end-point \textsf{D} of SU(3) QCD with dynamical quarks. 
Since center symmetry is explicitly broken by dynamical quarks \footnote{For an attempt to define an order parameter in the presence of dynamical quarks, see \cite{Fukushima}.}, the Polyakov loop acquires a nonzero expectation value $\langle L \rangle_{\textsf{D}}$ at \textsf{D}.
Thus the phase transition at \textsf{D} can be characterized by the infinite correlation length (or the vanishing screening mass) of the shifted Polyakov loop
\begin{eqnarray}
L'(\boldsymbol{x}) \equiv L(\boldsymbol{x})-\langle L \rangle_{\textsf{D}}. \label{l}
\end{eqnarray}
Using $L'$ as an order parameter,
 one can construct a Ginzburg-Landau theory near \textsf{D} and discuss the singular behavior of various susceptibilities.
However, due to our identification
\begin{eqnarray}
G(\boldsymbol{x}) \sim |L(\boldsymbol{x})|^2=|L'(\boldsymbol{x})+\langle L \rangle_{\textsf{D}} |^2,
\end{eqnarray}
the $0^{++}$ glueball field $G(\boldsymbol{x})$ contains a linear term in $L'$. Hence the screening mass of the Polyakov loop which vanishes at \textsf{D} is equal to the screening mass of the glueball, and we may as well construct a Ginzburg-Landau theory of $G$ around \textsf{D}. Note that this is a direct consequence of the explicit breaking of center symmetry. Unlike the SU(2) case, the glueball qualifies as an order parameter field; its susceptibility diverges with the exponent $\gamma$. 

The two descriptions of the critical phenomena at \textsf{D}, with the Polyakov loop or with the glueball, are equivalent. In the presence of dynamical quarks, the singular behavior of the Polyakov loop represents the character change of the glueball. Since the latter admits a direct physical interpretation, we will exclusively focus on the glueball in the following.

\subsection{Step 2: Gluball-sigma mixing}
The next step is a rather trivial, but important observation.
{\it In the presence of dynamical quarks, the glueballs must mix with the $\bar{q}q$ states}. In fact, the correct massless mode associated with the second order phase transition at \textsf{D} is a linear combination of the
pure glueball field $G$ and the scalar meson field $\sigma$ with large glueball content,
\begin{eqnarray}
\phi=G\cos \theta  + \sigma\sin \theta,\ \ \ \ \  \sin \theta \ll 1.\label{g}
\end{eqnarray}
 The other linear
combination with large sigma meson content,
\begin{eqnarray}
\phi'=-G\sin \theta + \sigma \cos \theta,
\end{eqnarray}
is massive, hence decouples at \textsf{D}.

The mixing Eq.~(\ref{g}) is physically obvious, and it is again a consequence of the explicit breaking of center symmetry. In the language of critical phenomena, the Ginzburg-Landau potantial has the following expansion,
\begin{eqnarray}
\Omega=[a(T-T_{\rm D})+b(m-m_{\rm D})]G'+O(G'^2)+O(G'^3)+O(G'^4), 
\end{eqnarray}
where ($T_{\rm D}$, $m_{\rm D}$) is the location of \textsf{D}, and $a$ and $b$ are smooth functions of $T$ and $m$. $G'$ is an interpolating field of the glueball whose expectation value at \textsf{D} is subtracted \footnote{One can also expand $\Omega$ in powers of $L'$. The coefficient of the linear term is unchanged up to a multiplicative constant.}.
The equation of the tangential line at \textsf{D} is given by
\begin{eqnarray}
a(T-T_{\rm D})+b(m-m_{\rm D})=0,
\end{eqnarray}
which are in general not parallel ($b\neq 0$) to the mass axis in the absence of symmetry.
The linear dependence on $m$ in the coefficient of $L'$ makes the chiral susceptibility
\begin{eqnarray}
\chi_{\rm ch} \sim  \int {\rm d}^d x\langle \bar{q}q(\boldsymbol{x})\bar{q}q(\boldsymbol{0})\rangle,
\end{eqnarray}
divergent with the same exponent $\gamma \sim 1.2$ as the Polyakov loop susceptibility, the glueball field susceptibility and the specific heat \footnote{However, if the tangential line is almost parallel to the mass axis, i.e., if $-\frac{a}{b}$ is very large, the mixing is weak and it would be difficult to observe the singularity of $\chi_{\rm ch}$ in numerical simulations \cite{fuku2}.}.

In the same way, near \textsf{C} the glueball field mixes with the sigma meson field which is critical. The simultaneous enhancement of the chiral susceptibility
and the Polyakov loop susceptibility along {\it each} crossover phase transition line is a natural consequence of the mixing \cite{satz,fuku}.

\subsection{Step 3: Level repulsion}

The coincidence of the peak positions is trivial. It is not a puzzle. It is just an evidence of the glueball-sigma mixing. The real puzzle of the lattice data is that there is only {\it one} such coincidence in the $(T, m)$ plane. Namely, the mixing Eq.~(\ref{g}) alone does not exclude the phase diagram Fig.~\ref{2} in which the coincidence occurs {\it twice}, first at the deconfinement phase transition and then at the chiral phase transition. So what is going on in Fig.~\ref{fig}?

The last step of our argument is a possible bridge between the two phase transitions.

We have defined the $\phi$ field Eq.~(\ref{g}) as the correct massless field at the end-point of the deconfinement phase transition \textsf{D}. On the other hand, since the massive $\phi'$ field has a large meson component there, naively we expect that it becomes critical (massless) at the end-point of the chiral phase transition \textsf{C}. This is what has been implicitly assumed to occur.
%
%
However, this innocent-looking assumption is the principal cause of our confusion. The two phase transitions are caused by {\it different} fields. Why do the two crossover lines meet?  The crucial difference between Fig.~\ref{2} and Fig.~\ref{fig} is that in Fig.~\ref{fig}, {\it both phase transitions are caused by the $\phi$ field}. Since the critical field at \textsf{C} should be predominantly the sigma meson, this is possible only if 
 the mixing angle changes from $\theta \sim 0$ to $\theta
\sim \frac{\pi}{2}$ in an intermediate region of the quark mass. 
 Such a continuous variation of the mixing angle is typical of a {\it level repulsion} in quantum mechanics. Thus we have reached a novel scenario of the finite temperature QCD phase transition: {\it Due to a level repulsion between the $\phi$ and the $\phi'$ fields, the $\phi$ field continues to be the lightest field for all values of the quark mass. Simultaneous divergences and peaks in various susceptibilities are simply caused by the dropping of the $\phi$ field screening mass.} This gives a natural explanation to the otherwise puzzling lattice result, Fig.~\ref{fig}. 

Note that the somewhat counterintuitive ordering 280 MeV $>$ 170 MeV mentioned in Section II is also a natural consequence of our scenario. Generally, a crossover line extends to the direction opposite to the symmetry broken phase. If the phase transition is second order in the absense of external fields, even the equation of the crossover line is determined by the universality class up to a multiplicative constant. The point is that center symmetry is broken in the {\it high} temperature phase. This is a quite unusual phenomenon but is true for SU($N_{\rm c}$) pure gauge theories. Thus, if the two phase crossover lines are smoothly connected by our scenario, the critical temperature of the deconfinement phase transition in the pure gauge theory must be larger than that of the chiral phase transition in the chiral limit.\\

\begin{figure}
\begin{center}
\includegraphics[scale=0.7]{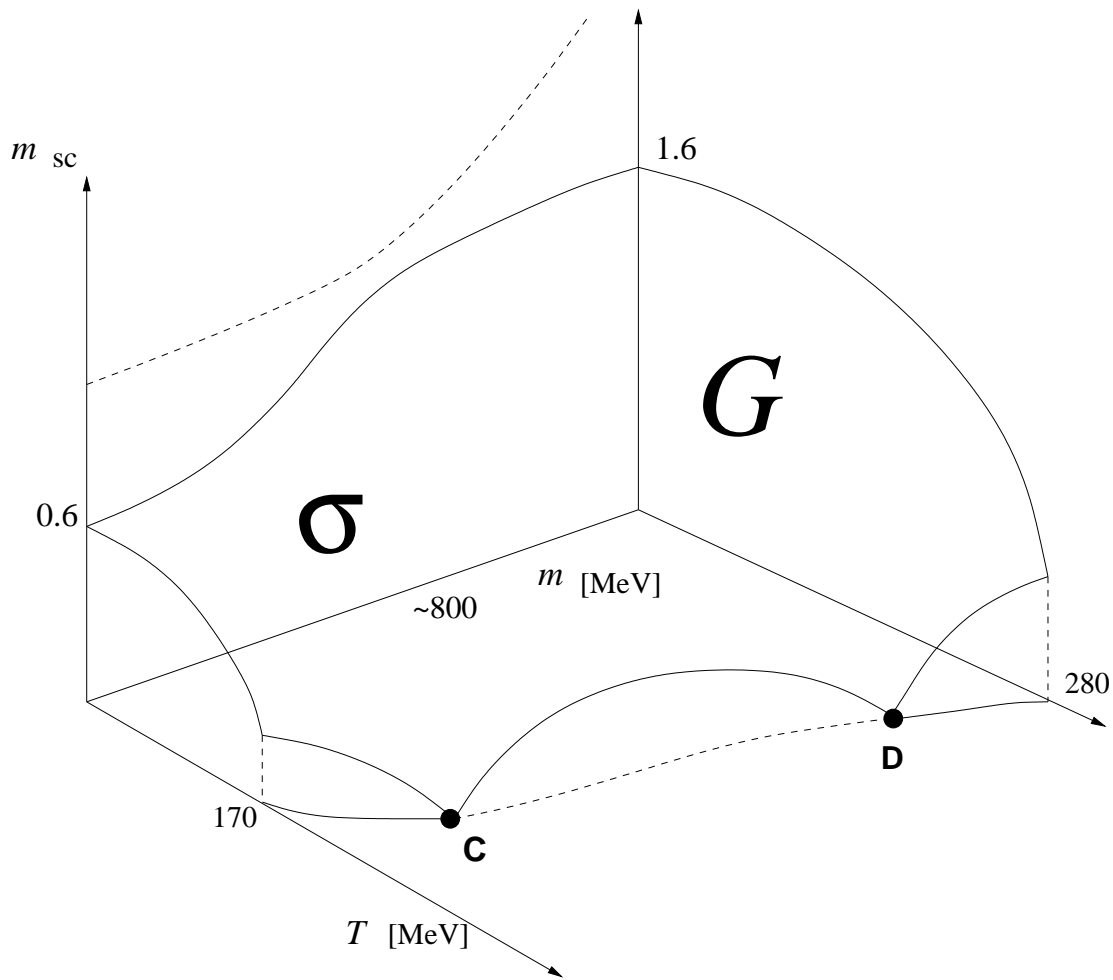}
\end{center}
  \caption{\label{phase} The lowest screening mass $m_{\rm sc}$ [GeV] in the scalar channel below the crossover phase transition. At small $m$, the scalar channel is dominated by the sigma meson $\sigma$. At large $m$, it is dominated by the glueball $G$.}
\end{figure}

We now argue that our level repulsion scenario is not an {\it ad hoc} idea just to explain the observed data, but rather a global feature of the QCD phase diagram in the ($T$, $m$) plane. Consider {\it zero} temperature QCD. For small values of the quark mass, the sigma meson is a broad resonance whose mass is about 600 MeV. On the other hand, the mass of the lowest glueball state (which mixes with nearby quarkonium states) is about 1.6 GeV.
If the quark mass is sent to infinity, the sigma meson mass goes to infinity, while the glueball mass stays more or less the same. Naively, this implies that there is a level crossing between the two states at an intermediate value, roughly half the glueball mass $\sim 800$ MeV, of the quark mass. 
However, in a quantum mechanical system, quite generally levels with the same quantum numbers cannot cross unless certain matrix elements accidentally vanish. Therefore, what actually takes place is a level repulsion. Under this assumption at $T=0$ (which seems to be reasonable to us), the level repulsion would continue to take place in the ($T$, $m$) plane up to the crossover phase transition 
and our scenario would be realized \footnote{If the $\phi'$ field were the order parameter at \textsf{C}, there must have been a level crossing between the $\phi$ and $\phi'$ fields. Generally, in a two-parameter space ($T$, $m$), a level crossing occurs along a line simply because of a geometrical reason \cite{landau}. Therefore, this (conventional) scenario would lead to a level crossing at zero temperature, which is again unlikely.}.
Thus the glueball-sigma mixing and the level repulsion provide not only a unified description of the chiral and deconfinement phase transitions, but also a consistent characterization of the entire QCD phase diagram in the ($T$, $m$) plane.
 Fig.~\ref{phase} is a schematic plot of the lowest scalar screening mass below the crossover phase transition in our scenario. The two phase transitions are indistinguishable. Roughly speaking, they are the same phase transition.
At about $m \sim $ 800 MeV, there is a smooth change of the degrees of freedom, i.e., mesons at smaller $m$ to glueballs at larger $m$, in the low energy effective theory of QCD at {\it any} temperature at least up to the crossover phase transition.
In Section IV, we will see another possible example of the level repulsion.\\

\section{related theories}

\subsection{Finite density}
 Basically our argument does not change at nonzero baryon/isospin chemical potential.
The interplay between the sigma and the glueball (or the Polyakov loop) continues to exist \cite{allton,toublan,fuku2,fodor}. The only new feature is that as the QCD critical point is approached \footnote{We temporarily fix the quark masses such that the critical point exists at finite density.}, the baryon number density begins to mix with the scalar mode and the baryon number susceptibility diverges with the same critical exponent as the chiral susceptibility \cite{misha,hatta,fujii}.

\subsection{QCD with adjoint quarks}
Lattice simulations \cite{kogut2,karsch3} as well as model calculations \cite{fuku,sa} have shown that, in QCD with adjoint quarks, the chiral phase transition occurs at a distinctly higher temperature than the deconfinement phase transition. Here we analyze this problem from our point of view. For simplicity, we work in $N_{\rm c}=2$. Since adjoint quarks do not break center symmetry, the second order deconfinement phase transition persists for all values of the quark mass. The effect of the adjoint quarks is simply to modify the coefficients of the Ginzburg-Landau potential $\Omega$. Near a critical point ($T_{\rm c}$, $m_{\rm c}$)  on the critical line, $\Omega$ can be expanded as
\begin{eqnarray}
\Omega=[a(T-T_{\rm c})+b(m-m_{\rm c})]L^2+cL^4,\label{omega}
\end{eqnarray}
where $a$, $b$ and $c$ are smooth functions of $T$ and $m$.
From (\ref{omega}) we see that both the chiral susceptibility and the specific heat diverge with the critical exponent $\alpha$. This means that although the glueball and the sigma meson are massless on the critical line, they are distinct from the Polyakov loop whose susceptibility diverges with the exponent $\gamma$. The mixing between the Polyakov loop and the sigma meson is prohibited by center symmetry. On the other hand, at the chiral phase transition the chiral susceptibility diverges with the exponent $\gamma$ \footnote{The value of $\gamma$ may be different from that at the deconfinement phase transition depending on the universality class.}. Clearly, the two phase transitions cannot be simultaneous.
Interestingly, even in this case there is a level repulsion (or a level crossing) between the two phase transitions. Since the critical line is almost parallel to the quark mass axis ($b\approx 0$) for small $m$, the amplitude of the singular part of the chiral susceptibility is small. This means that the massless mode responsible for the $\alpha$-divergence is predominantly the glueball.  On the other hand, the massless mode at the chiral phase transition is predominantly the sigma meson. Naturally, we expect that there is a level repulsion  between the sigma meson and the glueball at an intermediate temperature.

\subsection{$N_{\rm c}\neq 3$}
It is intriguing to consider the phase diagram for different values of $N_{\rm c}$ than three and in the limit of infinite $N_{\rm c}$ \cite{thorn,neri,pisarski,karsch4}.
Lattice simulations have shown that, as $N_{\rm c}$ increases, the first order deconfinement phase transition strengthens \cite{teper}. This means that a smaller value of $m$ is needed to wash out the first order phase transition, i.e., the point \textsf{D} moves downward.  Since there does not seem to be anything special about three colors in our scenario, we expect that the point \textsf{D} {\it reaches} the point \textsf{C} at some value of $N_{\rm c}=N_{\rm c}^*$ and the high and low temperature phases are separated by a single first order phase transition line for $N_{\rm c}\ge N_{\rm c}^*$. 

However, the limit $N_{\rm c} \to \infty$ might be a different story. The critical temperature of the deconfinement phase transition is expected to be of the order of $\Lambda_{\rm QCD} \sim O(1)$, while that of the chiral phase transition is controversial. If we regard $4\pi f_{\pi}$ as the chiral symmetry breaking scale ($f_{\pi}$ is the pion decay constant), it is of the order of $f_{\pi} \sim O(\sqrt{N_{\rm c}})$. 
Our scenario appears to be incompatible with this scenario \footnote{It is possible that the two phase transitions are connected even in this case. However, we regard this possibility as unnatural in view of our remarks in the previous section and dismiss it.}. On the other hand, there are arguments that the chiral phase transition temperature cannot grow with $N_{\rm c}$; it is of the order of the $\rho$ meson mass $m_{\rho} \sim O(1)$.

 It is not our intention to go into the details of these arguments which have been considered by many people previously. Here we just point out that at least our scenario breaks down in the large-$N_{\rm c}$ limit. This is because, in the usual power counting rule of large-$N_{\rm c}$ QCD, the glueball-meson mixing strength vanishes like $\sim 1/\sqrt{N_{\rm c}}$. In the absence of the mixing, the level repulsion is impossible and the `unlocking' of the two phase transitions  could occur. 

From our viewpoint, $N_{\rm c}=2$ is special. The deconfinement phase transition in the pure gauge theory is second order. The Ginzburg-Landau theory of $G$ does not make sense at \textsf{D}. Nevertheless, the mixing between the Polyakov loop and the glueball is possible when the dynamical quarks are added.
Another notable feature is that the phase transition on the `quark side' is driven by the $0^+$ diquark (`baryon') at finite temperature and density. It is thus interesting to pursue a similar link (glueball-sigma-diquark mixing) in the phase diagrams of two-color QCD \cite{nishida}.

\section{Conclusions and outlook}

Based on the existing lattice data, we have proposed a novel scenario of the finite temperature QCD phase transition. The chiral phase transition at zero quark mass and the deconfinement phase transition at infinite quark mass are continuously connected by the glueball-sigma mixing and the level repulsion.
 This gives a precise meaning to the common but somewhat ambiguous statement: ``In QCD at high temperature, quarks are deconfined and chiral symmetry is restored." 

We have also argued that our scenario can be naturally embedded in the entire phase diagram in the ($T$, $m$) plane (Fig.~\ref{phase}).
At any given value of the quark mass, the minimum of the $\phi$ field screening mass marks the well-defined onset of the QGP phase.
\\

Although we believe that our scenario is the correct interpretation of the finite temperature lattice simulations, direct evidences are still lacking. The most convincing evidence would be obtained by the lattice QCD measurements of the glueball, the Polyakov loop and the sigma meson screening masses along the crossover transition line. Because of the mixing Eq.~(\ref{g}), they are all equal along the line. 
In principle, the mixing between the sigma and the glueball could occur in the entire phase diagram (Fig.~\ref{fig}). However, we expect that in most regions the mixing is too small to be observed. For example, well above the chiral phase transition temperature, for small $m$, the sigma meson and the pions fall in a multiplet due to the restored chiral symmetry. The glueball mass is indeed different from the sigma mass in this region \cite{gavai}. Also at low temperatures, for small $m$, the glueball will primarily mix with nearby scalar quarkonium states. 
As argued in Section III, the strong mixing is expected to occur at rather large values of $m$, say, $m\sim 800$ MeV. Detailed analyses of the lowest and the second lowest screening masses in this region would confirm or exclude our scenario.

For a deeper understanding of this phenomenon, it would be desirable to construct  concrete models which embody the mixing mechanism. In models where the Polyakov loop and the sigma are coupled in the effective potential, the pseudocritical temperatures of the two phase transitions are more or less close to each other \cite{gock,fuku,fuku2}. However, it has been so far not successful to reproduce the complete `locking' of the two phase transitions. The obtained phase diagram is qualitatively similar to Fig.~\ref{2}, which suggests that although these models capture the gross features of the phase transition, an essential coupling which reproduces the lattice data and demonstrates our scenario is still missing. This appears to be a future theoretical challenge.

{\bf Acknowledgements:}

We thank S.~R.~Beane, M.~Creutz, S.~Ejiri, L.~McLerran, P.~Petreczky, R.~D.~Pisarski, D.~T.~Son, K.~Splittorff and M.~A.~Stephanov for discussions and comments.
Y.~H.\ is supported by RIKEN BNL Research Center and U.S. Department of Energy grant \#DE-AC02-98CH10886. K.~F.\ is supported by the Japan Society for the Promotion of Science for Young Scientists and supported in part by funds provided by the U.S. Department of Energy under cooperative research agreement \#DF-FC02-94ER40818.

\end{document}